\documentclass[a4paper]{article}

\usepackage[pdftex]{graphicx}
\usepackage{lipsum,amsmath,multicol}

\usepackage{latexsym}
\usepackage{dcolumn}
\usepackage{amsmath}
\usepackage{epsf}
\usepackage{float}
\usepackage{hyperref}

\newcommand{\x}{\mathrm{x}}

\newcommand{\tr}{\text{tr}}

\def\W{\mathrm{W}}

\def\M{\mathcal{M}}

\def\E{\mathcal{E}}
\def\N{\mathcal{N}}

\def\tr{\mbox{tr}}
\def\bea{\begin{eqnarray}}
\def\eea{\end{eqnarray}}

\begin{document}
\title{Distilling Entanglement with Noisy Operations}

\author{Jinho Chang$^{1}$, Joonwoo Bae$^{2}$, and Younghun Kwon$^{1}$ \\
%Physical Review Letters.)}\\
\small{Departments of Applied Physics$^{1}$ and Applied Mathematics$^{2}$,} \\
\small{ Hanyang University (ERICA), 55 Hanyangdaehak-ro, Ansan,} \\
\small{Gyeonggi-do, 426-791, Korea.}  \\
%$^{2}$ Department of Applied Mathematics, Hanyang University (ERICA), 55 Hanyangdaehak-ro, Ansan, Gyeonggi-do, 426-791, Korea 
}

 \maketitle

\begin{abstract}
Entanglement distillation is a fundamental task in quantum information processing. It not only extracts entanglement out of corrupted systems but also leads to protecting systems of interest against intervention with environment. In this work, we consider a realistic scenario of entanglement distillation where noisy quantum operations are applied. In particular, the two-way distillation protocol that tolerates the highest error rate is considered. We show that among all types of noise there are only four equivalence classes according to the distillability condition. Since the four classes are connected by local unitary transformations, our results can be used to improve entanglement distillability in practice when entanglement distillation is performed in a realistic setting. 
\end{abstract}

%protocol and its application to long-distance communication with quantum repeaters, assuming that channels and measurement might not be stabilized so that they can remain ever noisy whatsoever. 

%\begin{IEEEkeywords}

% \fi
%\IEEEpeerreviewmaketitle

\section{Introduction}
 
Entanglement distillation, the art of distilling entanglement of pure states, poses one of the fundamental tasks for quantum information applications \cite{ref:IBM} \cite{ref:Oxford} \cite{ref:3} \cite{ref:4}. The purification process not only extracts entanglement but also gets rid of any effect of decoherence. Consequently, it protects systems of interest against intervention of environment. This  has a number of applications. For instance, entanglement distillation is a key technique to extend the communication distance \cite{ref:Briegel} \cite{ref:Dur}. Contrast to classical signals, quantum states cannot be cloned \cite{ref:noclone}, that is, they cannot be amplified, and the distance that quantum states can be shared is limited. It turns out that entanglement distillation and entanglement swapping lead to achieving long-distance quantum communication \cite{ref:Briegel}. 
 
%Entanglement distillation bears some similarity with secret key distillation of symmetric cryptography. To distill entanglement one has to build a protocol that applies local operations and classical communication (LOCC). For instance, one-way error correction protocols can be applied to distilling entanglement via one-way communication. The two-way communication protocol called advantage distillation in symmetric cryptography \cite{ref:mau} has been adapted to entanglement distillation \cite{ref:IBM} \cite{ref:Oxford}, which we call the two-way distillation protocol throughout. With the protocol at hand, it has been shown that all two-qubit entangled states are distillable. For high-dimensional quantum systems, the condition of entanglement distillability remains open - currently the condition is only introduced by referring to specific protocols \cite{ref:IBM} \cite{ref:Oxford} \cite{ref:durcirac} \cite{ref:high}. 

%There is some similarity between entanglement of quantum states and secret correlations in probabilities. 

Distillable entangled states have been characterized by referring to specific distillation protocols. The distillation protocol that tolerates highest error rates has been obtained by adapting a secret key distillation protocol \cite{ref:mau} to entanglement distillation \cite{ref:IBM} \cite{ref:Oxford}, which we call the two-way distillation protocol throughout. It has been shown that, by exploiting the two-way entanglement distillation protocol, all two-qubit entangled states are distillable.

A realistic scenario such that entanglement distillation is performed with noisy operations has been considered, when random noise corresponding to the depolarization channel appears \cite{ref:Dur}. This is a worst-case scenario consideration since a specific type of noise can be randomized by local operations and classical communication \cite{ref:Briegel} \cite{ref:Dur}. Then, it turns out that if noisy operations are applied in the protocol, two-qubit entangled states are no longer distillable in general. This would seek a possibility of improving entanglement distillation protocols in such a way that they are robust against noisy operations. It is also of practical interest to characterize those two-qubit states which are distillable when local operations are applied together with classical communication.

%For high-dimensional quantum systems, the condition of entanglement distillability remains open - currently the condition is only introduced by referring to specific protocols \cite{ref:IBM} \cite{ref:Oxford} \cite{ref:durcirac} \cite{ref:high}. 

%Therefore, it turns out that so far the two-way protocol tolerates the highest error rate. This is significant in long-distance communication where relays or repeaters involved in entanglement swapping generate high rates of errors. 
 
%That is, due to noise, the protocol will saturate at an entangled state with reduced fidelity. The strategy of distilling entanglement is then to necessarily combine one-way protocols with the two-way protocol: the two-way protocol runs until it reaches the threshold that one-way protocols tolerate. All these are connected to related applications of entanglement distillation, such as long-distance communication. 

One can then naturally ask about cases where noise that appears is not random. For instance, realistic constraints in experiment can be identified and then properties of given devices in a laboratory can be found in advance. In such cases, one does not have to apply depolarization but can immediately restrict types of errors that may appear. Learning the properties, one can compare two cases of, i) randomizing types of noise, that is, depolarization, and ii) not randomizing but keeping particular types of noise. The analysis to the latter is lacking.

%\subsubsection*{Summary of Results} 

In the present work, we investigate the two-way distillation protocol with noisy quantum operations. We show that, in terms of the distillability condition, all types of noise are in fact grouped to only four equivalence classes, i.e., any pair of types of noise in the same class are equivalent in terms of the distillability condition. Their specific forms are to be presented later and the four inequivalent groups can be summarized in brief. 
\begin{itemize}
\item Error Class (I) commutes with the protocol and, hence, has no effect to the distillability condition. When these errors are found in the characterization of noisy quantum operations, one safely concludes that they are not affecting to the distillability condition. 
\item Error Class (M) denotes cases that noise appears in measurement asymmetrically. The distillation protocol is more robust to (M) than the depolarization. 
\item Error Class ($C_1$) denotes a set of types of noise in channels: the distillation protocol works worse under ($C_1$) than the depolarization. 
\item Error Class ($C_2$) denotes a set of the other type of noise in channels: the distillation protocol works worst under the types of noise in this class.  
\end{itemize}
The depolarization noise, or equivalently random noise, then corresponds to the case that all types of noise in the above appear with equal probabilities. As it is motivated in the beginning, the results show that there are a number of types of noise to which the protocol is less affected than the depolarization channel, such as classes (I) and (M). Note also that classes in the above are transformed by local unitary transformations. Hence, once experimental conditions are found in $(C_1)$ or $(C_2)$, there exist local unitary transformations such that the experimental conditions are manipulated to be in (I) or (M). Our results can be used to devise an entanglement distillation protocol that would be more resilient to local noise. The results can also be applied to extending the communication distance, compared to the previous consideration of depolarization channels \cite{ref:Briegel}. 

This paper is structured as follows. We firstly review the distillation protocol in the setting of ideal and depolarization noise. We present a formalism of considering noise model and introduce a noisy distillation map. We then apply our general formalism and analyze consequences of noise effects. This classifies various types of local noise into only four classes.

\section{The entanglement distillation protocol under depolarization noise}

In this section, we briefly review the two-way entanglement distillation protocol and its performance under a depolarization noise and measurement imperfections.

\subsection{The protocol}

There are two protocols of entanglement distillation with two-way communication \cite{ref:IBM} \cite{ref:Oxford}. Although having different efficiency, they are actually equivalent in terms the distillability condition. As we here focus on distillability, we consider the protocol proposed in \cite{ref:IBM} for convenience, that keeps Werner states during runs of the protocol in which entanglement properties are completely found by a single parameter. 

To begin the protocol, let us suppose that two parties share $N$ copies of two-qubit states. The goal of entanglement distillation is to transform them into a number of maximally entangled states by local operations and classical communication. A single run of the two-way distillation protocol is composed of three steps in the following. 

\subsubsection{Twirling operation} The first task to do is to apply so-called the twirling operation. It is also a protocol that can be implemented with local operations and classical communication, and transforms shared pairs of quantum states into Werner states, 
\bea
\rho_{W} (F) = F  |\phi^{+} \rangle \langle \phi^{+}| + \frac{1-F}{3} (I - |\phi^{+}\rangle \langle \phi^{+}|), \label{eq:wstate}
\eea
where $|\phi^{+}\rangle = ( |00\rangle +  |11\rangle )/\sqrt{2}$ is the maximally entangled state and is to be distilled at the end of the protocol. The entanglement property of Werner states can be completely characterized by a single parameter $F$, called the singlet fidelity $F = \langle \phi^{+} |  \rho_{W} (F) | \phi^{+} \rangle $. Note that Werner states are entangled if and only if $F>1/2$ and otherwise, separable \cite{ref:werner} \cite{ref:PereHoro}.  

The twirling operations can be implemented using unitary transformations picked up according to the Haar measure, or in practice, performed by using a finite number of unitaries. These unitaries form the so-called $2$-design \cite{ref:2design}, denoted by $\{ U_{k} \}$, and can be applied as 
\bea
\mathcal{T}(\rho) = \sum_{k} U_k \otimes U_k \rho U_{k}^{\dagger} \otimes U_{k}^{\dagger} = \rho_{W} (F). \label{eq:twirl}
\eea
Note that the singlet fidelity given in shared states in the beginning does not change under the twirling operation, i.e. $F = \langle \phi^{+}|   \rho|\phi^{+}\rangle  = \langle \phi^{+} |   \rho_W |\phi^{+}\rangle $. The operation randomizes two-qubit states but the maximally entangled state.

\begin{figure}[h!]
\centering\rule{0pt}{4pt}\par
\scalebox{1}{\includegraphics[width= 8.5cm]{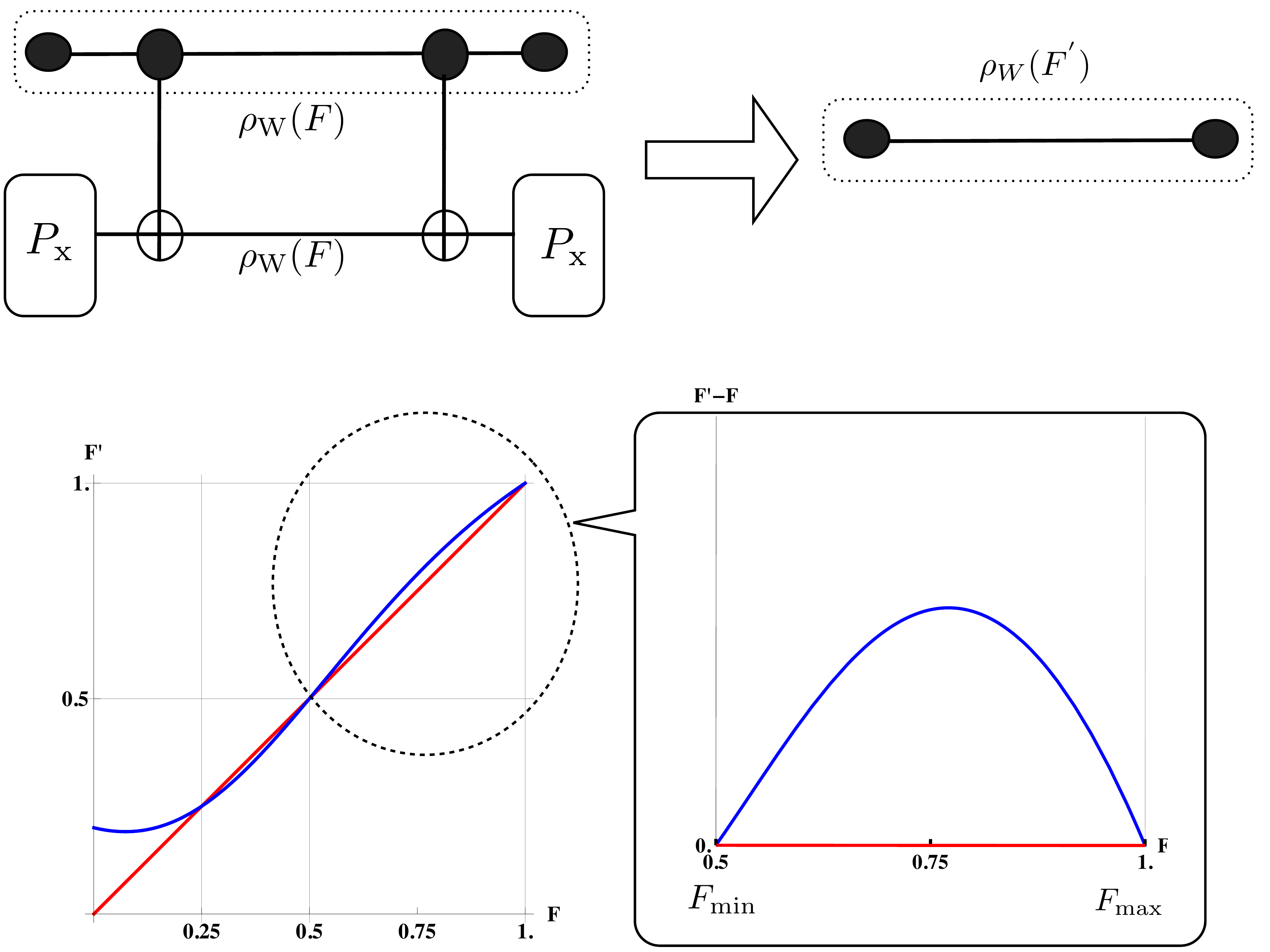}}
\vspace*{8pt}
\caption{\label{fig:fig1}
The two-way entanglement distillation protocol is comprised of the bilateral CNOT operation on two copies of Werner states characterized by fidelity $F$ and measurement on the second one. The projective measurement in the second register is denoted by $P_{\x} = | \x \rangle \langle \x |$ for $\x=0,1$. The resulting pair in the first register is accepted only when the measurement outcomes are equal. Denoted by $F^{'}$ the fidelity of a resulting state in the first register, the relation between $F$ and $F^{'}$ is then shown that $F^{'} > F$ if $F>1/2$. This shows that as the protocol runs, the singlet fidelity converges to the unit: hence, maximally entangled states are distilled by running the protocol. }
%We write by $F_{\min}$ and $F_{\max}$, minimal and maximal fidelities respectively between which entanglement can be distilled.  In the entanglement distillation using ideal quantum operations, all entangled Werner states, i.e. for $1/2=F_{\min}<F\leq F_{\max}=1$, can be distilled. }
\end{figure}

\subsubsection{Bilateral CNOT} Once Werner states are shared between the parties, the next is to perform the bilateral controlled-NOT (CNOT) operation, see also Fig. \ref{fig:fig1}. Taking two copies of shared Werner states, denoted as $A_1 B_1 A_2 B_2$, it applies the CNOT operations on parties $A_1 A_2$ and $B_1 B_2$ respectively.  This works by pairing two copies of Werner states: that is, we have
\bea
| a_1\rangle_{A_1}| a_2\rangle_{A_2} & {\longrightarrow} & | a_1\rangle_{A_1} |a_1 + a_2\rangle_{A_2} \nonumber \\
| b_1\rangle_{B_1}| b_2\rangle_{B_2} & \longrightarrow & | b_1\rangle_{B_1} |b_1 + b_2\rangle_{B_2} \nonumber 
\eea
Let $\E_{U}$ denote the bilateral operation over two copies of Werner states.

\subsubsection{Post-selection} Having done the bilateral CNOT operation, projective measurement in the computational basis is applied to the second register. If measurement outcomes of the second register in both sides are equal, the remaining state in the first register is accepted. Otherwise, all are discarded and the protocol repeats to other copies. We write by $F^{'}$ the singlet fidelity of a resulting state once the first register is accepted.

\subsubsection{Distillability} The entanglement property of Werner states is completely characterized in terms of the singlet fidelity. The maximally entangled state corresponds to the case that $F=1$. The whole process of entanglement distillation can be understood as increasing the singlet fidelity from a lower value to the unit: given singlet fidelity $F$, entanglement can be distilled by repeating a protocol if the fidelity $F'$ resulting from the protocol is larger than the initial one $F$. Let us define the singlet fidelity increment $\delta F$ as
\bea
\mathrm{singlet}~ \mathrm{fidleity} ~\mathrm{increment}: ~~~\delta F = F' - F. \label{eq:fincre}
\eea
In summary, distillability is now equivalent to whether $\delta F$ is positive, or not. In fact, it turns out that entanglement can be distilled from all two-qubit entangled states using the protocol. That is, as long as $F>1/2$, we have $\delta F >0$ after a round of the protocol, see Fig. \ref{fig:fig1}. 

%To derive this, let us begin with recalling that, using stochastic LOCC,  all two-qubit states can be transformed into Werner states without losing any entanglement properties. That is, two-qubit states can be filtered into entangled Werner states if and only if they are entangled. Once this is done, it holds that $\delta F >0$ if and only if $F>1/2$. Thus, in this way, it can be shown that all two-qubit entangled states are distillable.

\subsection{Distillation under depolarization noise}

In a realistic setting it is natural to consider operations in the entanglement distillation protocol are not ideal but noisy due to interaction with environment. Recall that the operations contain a collective operation over two copies, bilateral CNOT denoted by $\E_{U}$ over $A_1 A_2$ and $B_1 B_2$ respectivley, and the other, projective measurement in the second registers $A_2$ and $B_2$. Note that noise appearing through channels, as well as noise in the twirling operations, are all included in the Werner states shared between two parties. This means that identification of the singlet fidelity $F$ of initially given Werner states takes all of noise effects right before the protocol, into account. 

The two-way entanglement distillation with noisy operations has been considered in \cite{ref:Dur} \cite{ref:geza}. Two cases are considered: firstly, it is assumed that noisy operations may appear randomly, and secondly, measurement in the second register contains imperfections.  

%Then, from results of the measurement, two parties share a quantum state in the first register with some probability, and the singlet fidelity of the resulting state is inferred from both of the success of the protocol and of an initially given fidelity $F$. Therefore, considering noisy operations can be done by looking at cases when noise happens in channels or in measurement.

%\label{ssbs:complete-depolarization}

\subsubsection{Random noise to the bilateral operation} 

Recall that the bilateral CNOT operation is denoted by $\E_{U}$, and suppose that some of the four locations $A_1 A_2 B_1 B_2$ that apply the operations are coupled to environment, hence where noisy operations are performed. In \cite{ref:Dur}, it is supposed that locations of local noise are not known, and therefore more generally as the worst consideration that random noise has happened  has been investigated. This is equivalent to the presence of depolarization noise to the operation $\E_{U}$. Let $p$ denote the probability random noise happens to $\E_{U}$, and then the resulting state after the bilateral CNOT operation $\E_{U}$ is expressed as follows,
\bea
\E_{U} (\rho_{W}^{\otimes 2} (F)) \rightarrow (1-p) \E_{U}(\rho_{W}^{\otimes 2}(F)) + p \frac{\mathrm{I}_2^{\otimes 4} }{16}, \label{eq:deo}
\eea
where $\mathrm{I}_2$ denotes the identity in the two-dimensional Hilbert space. Measurement in the computational basis is applied in the second register. Note that one can always restrict the consideration to the case since any type of local noise in some of the four locations can be mapped to a depolarization by a randomization process.

\begin{figure}[h!]
\centering\rule{0pt}{4pt}\par
\scalebox{1}{\includegraphics[width= 9cm]{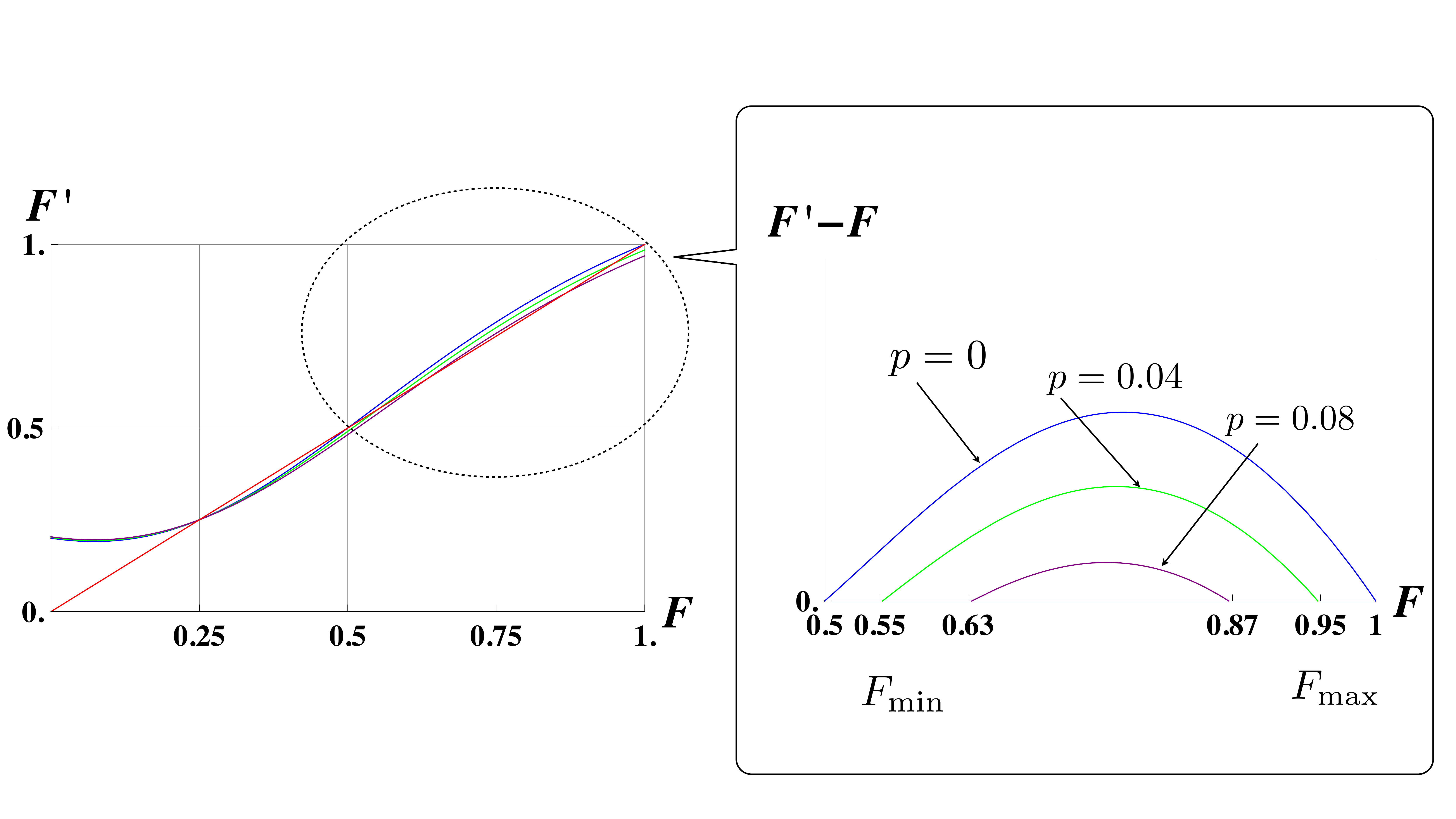}}
\vspace*{8pt}
\caption{\label{fig:fig3chno}
Entanglement distillation when a probabilistic depolarization channel is present is shown. The ideal corresponds to the case $p=0$, see the noise parameter explained in Eq. (\ref{eq:deo}). It is observed that two distillability parameters $F_{\max}$ and $F_{\min}$ no longer span all entangled states; thus, all entangled states are not distillable any more. Entanglement can be distilled from Werner states $\rho_{W}(F)$ only when $F>F_{\min}$.}
\end{figure}

It has been shown that due to the noise, the condition of entanglement distillability is significantly modified. In particular, all entangled two-qubit states are no longer distillable: i.e., from Eq. (\ref{eq:fincre}, it does not hold any more $\delta F>0$ for all $F>1/2$. To be precise, we introduce two parameters of distillability, $F_{\min}$ above which the protocol increases the fidelity i.e. $\delta F>0$ and $F_{\max}$ above which the protocol no longer work $\delta F=0$. For instance, we have $F_{\min}=1/2$ and $F_{\max}=1$ in the noise-free case. It has been found that by a depolarization noise the protocol works for $[F_{\min}, F_{\max}]$ with $F_{\min}>1/2$ and $F_{\max}<1$ \cite{ref:Dur}. In Fig. (\ref{fig:fig3chno}), the numerical result is reproduced for $p=0.04$, for which we have $F_{\min} =0.55$ and $ F_{\max} =0.95$, and also for $p=0.08$ for which we have $F_{\min} =0.63$ and $F_{\max} =0.87$. That is, the protocol under a depolarization noise works only for sufficiently entangled states and moreover cannot reach the maximally entangled states but Werner states with $F_{\max}$ at most. In this case, the two-way protocol runs until resulting singlet fidelity reach where one-way distillation protocols would work.

%This error can be considered as the case that two parties are indifferent to specific or particular types of errors as all possible cases of noise are randomized, i.e. depolarized with some probability. Assuming that the projective measurement in Eq. (\ref{eq:proj}) is perfect, the fidelity after a single iteration of the entanglement distillation protocol becomes accordingly, compared to the fidelity in Eq. (\ref{eq:newf})

%After all, what noise matters to the protocol of distilling entanglement is reduced into the entanglement distillability. Let us recall, from Sec. (\ref{sssec:distillability}), that entanglement can be distilled from all entangled two-qubit states. This follows from the observation that, after an iteration of the protocol, the singlet fidelity $F^{'}$ is probabilistically amplified from two copies of states of singlet fidelity $F$, and the succeeded case can be deterministically selected from the measurement outcomes in the second register. Thus, the essential question about effects of noise lies at how the singlet fidelity $F^{'}$ after an iteration of the protocol is related to the noise parameter of channels. From the depolarization model in Eq. (\ref{eq:deo}), once measurement is successfully done in the second copy, the resulting fidelity is found as, $(1-p)F^{'} + p1/16$, where $F^{'} = \langle \phi^{+} |   \E_{\U}(\rho_{W}^{\otimes 2} (F)) | \phi^{+}\rangle$ is the singlet fidelity expected from the ideal case of channels.

%Important is the result that, by noise in the channel $\E_{\U}$, the set of distillable states is modified. 

\begin{figure}[h!]
\centering\rule{0pt}{4pt}\par
\scalebox{1}{\includegraphics[width= 9cm]{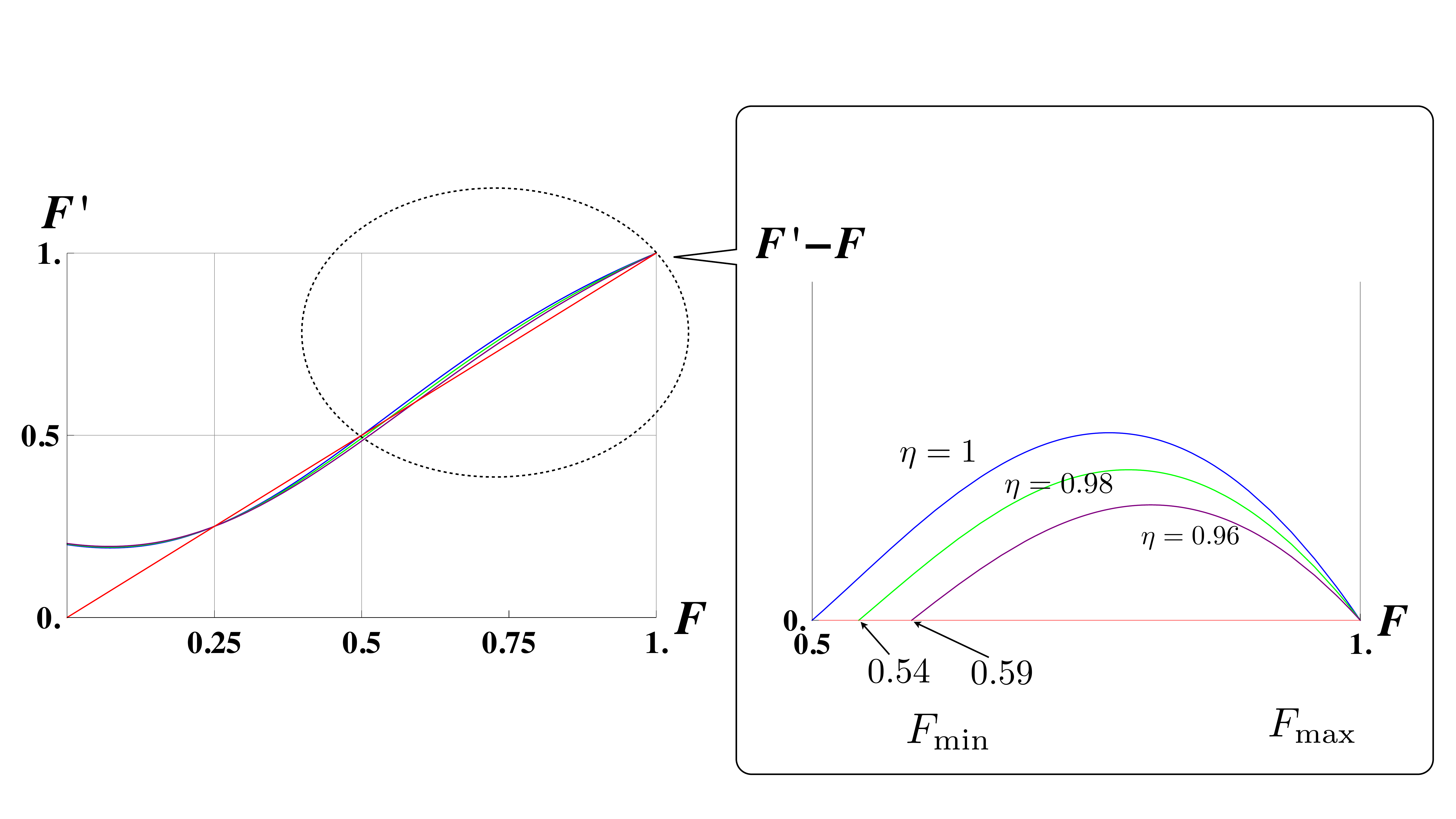}}
\vspace*{8pt}
\caption{\label{fig:fig4menon}
Entanglement distillation when the projective measurement is noisy is shown. It is assumed that measurement is equally noisy in the second register $A_2$ and $B_2$ in the protocol. The maximally attainable singlet fidelity is not modified i.e. $F_{\max}=1$, while the lowest threshold increases $F_{\min}>1/2$ depending on the noise parameter $\eta$ see Eq. (\ref{eq:dproj}).}
\end{figure}

\subsection{Identical imperfection in measurement}
\label{ssec:idenimpm}

Noise can also appear in measurement, i.e. in the projective measurement in the second register \cite{ref:Dur}. The measurement in the ideal case works with computational basis $\{ |0\rangle,|1\rangle\}$. Let $P_{\x} =|\x \rangle \langle \x |$ denote projective measurement for $x=0,1$. An \emph{imperfect measurement} can be described by,
\bea
\widetilde{P}_{\x} (\eta)  = \eta P_{\x} + (1-\eta)P_{\x +1}, ~~\mathrm{for}~\x=0,1 \label{eq:dproj}
\eea
where $\eta$ is the noise parameter and the addition is computed in modulo $2$. With imperfect measurement, the noisy measurement in the second register can be written as,
\bea
\widetilde{P}^{(A_2 B_2) }(\eta) = \widetilde{P}_{0}^{(A_2)} (\eta) \otimes \widetilde{P}_{0}^{(B_2)} (\eta) + \widetilde{P}_{1}^{(A_2)} (\eta) \otimes \widetilde{P}_{1}^{(B_2)} (\eta). \nonumber %\label{eq:dproj2}
\eea
Note that, in the above, the noise parameter $\eta$ is assumed to be equal to all measurement projections of Alice and Bob. 

With measurement imperfections in the second register, the range $[F_{\min} , F_{\max}]$ in which singlet fidelity increases, changes. Interestingly, while noise is present in measurement, it holds that $F_{\max} =1$. However, the lowest threshold does not cover all two-qubit entangled states: $F_{\min}>1/2$. In Fig. (\ref{fig:fig4menon}), a numerical result in \cite{ref:Dur} is shown for cases $\eta=0.98$ and $\eta=0.96$.

\section{Types of local noise in entanglement distillation}

So far, we have recalled the distillation protocol and its performance with noisy operations. In what follows, in the protocol we identify locations where local noise may happen and present detailed descriptions. We also show that measurement imperfections can be equivalently dealt as noise in operations.

\subsection{Local noise and the distillation map}
\label{sec:fomalism}

%and briefly review the known result with the depolarization noise. 

\subsubsection*{Pauli channel} We begin with fixing notations and terminologies on noisy quantum operations. Throughout, we write by $\N_{q}^{}$ a Pauli channel with the overall error rate $q$ and its composition to a qubit channel $\E$ works as follows, 
\bea
\N_{q} \circ \E (\cdot) = (1-q) \E (\cdot) + \sum_{i=x,y,z} r_i \sigma_i  \E  (\cdot) \sigma_i ,\label{eq:noisech}
\eea
where $\sigma_i$ are Pauli matrices, $i=x,y,z$. It is clear that $(1-q) + \sum_{i = x,y,z} r_i =1$ for $q \geq 0$ and $r_i \geq 0$. Note that once $r_i = r_j$ for all $i,j$, the Pauli channel becomes depolarization with probability $1-q$. One can also easily find that Pauli channels are self-dual:
\bea
\tr[ \N_q (\rho) A] & = & \tr[( (1-q)\rho + \sum_i r_i \sigma_i \rho \sigma_i   ) A ] \nonumber\\
%& = & \tr[(1-q)\rho A + \sum_i r_i \sigma_i \rho \sigma_i  A ] \nonumber\\
& = & \tr[  \rho (1-q) A + \sum_i r_i  \rho \sigma_i  A \sigma_i ] \nonumber\\
& = & \tr[  \rho \N_q (A)]. \label{eq:sedu}
\eea
This means that noise appearing in channels can be equivalently considered as noise appearing in measurement devices, and vice versa. This is to be exploited later to show the equivalence classes of types of noise.

\subsubsection*{The two-way distillation map} 

We here encapsulate bilateral CNOT and measurement in the same outcomes, which happen with probability, as the two-way distillation map that can be defined only with measurement outcomes are equal. Note that the map can be defined only when a round of the protocol is successful, i.e. measurement in the second register gives the same outcomes.

Denoted CNOT operation by $U_{CN}$, the bilateral CNOT gate over two copies in the local site $A_1 A_2$ can be expressed as, $U_{CN}^{(A_1 A_2)} = |0\rangle \langle 0| \otimes I  + |1\rangle  \langle 1 | \otimes \sigma_{x}$ where $\sigma_x$ is the Pauli $X$ matrix. Then, the bilateral CNOT operation $\E_{U}$ for two copies states $A_1A_2 B_1 B_2$ can be explicitly written as follows,
\bea
\E_{U}(\cdot) = U_{CN}^{(A_1 A_2)} \otimes U_{CN}^{(B_1 B_2)} ~(\cdot)~ U_{CN}^{(A_1 A_2)\dagger} \otimes U_{CN}^{(B_1 B_2)\dagger}. \label{eq:bc}
\eea
The next is measurement in the second register $A_2 B_2$ in the computational basis $P_{x} = |x \rangle \langle x|$ for $x=0,1$. Then, after measurement, two parties communicate each other and check if their measurement outcomes are equal or not. If they are equal, the shared state in the first register is accepted. Otherwise, they repeat the protocol with other copies. The post-processing is therefore probabilistic, and we define the following map $\E$ that describes the two-way distillation protocol in the case that the measurement outcomes in the second register are accepted as follows,
\bea
\E(\cdot) &  = &  p_{\mathrm{succ}}^{-1}  \tr_{A_2 B_2} [ \E_{U}(\cdot)~P^{(A_2 B_2)}  ],~\mathrm{where}~\label{eq:emap} \\
p_{\mathrm{succ}} & = & \tr_{AB} [ \E_{U}(\cdot)~P^{(A_2 B_2)}  ],~\mathrm{} \nonumber\\
P^{(A_2 B_2)} & = & P_{0}^{(A_2)} \otimes  P_{0}^{(B_2)} + P_{1}^{(A_2)} \otimes  P_{1}^{(B_2)}, \label{eq:projm}
\eea
and $AB$ denotes the four systems $A_1B_1 A_2 B_2$. We call the map $\E$ in the above the two-way distillation map that applies to Werner states and describes all local operations in the distillation protocol. The twirling operation is included in the process of preparing Werner states.

\subsection{Local noise in entanglement distillation}

\subsubsection*{Motivation and assumptions} 

Let us now introduce the noise model that we are going to consider in the entanglement distillation protocol. The basic assumption is that local devices are coupled to environment individually and locally such that quantum noise appears in single qubit operations. In general, the coupling means that system and environment $\rho_s \otimes \rho_{\mathrm{env}}$ evolves under a unitary $U_{S+E}$ together, and consequently the resulting state is often an entangled state of both systems, e.g. $|\Psi\rangle_{S+E} = \sum_k  \sqrt{p_k}| \psi_k\rangle |e_k\rangle$, in which system dynamics shows that, $\rho_s \rightarrow\tr_{\mathrm{env}} |\Psi\rangle_{S+E} \langle \Psi | = \sum_k p_k |\psi_k \rangle \langle \psi_k |$. 

For qubit operations, the noise map in Eq. (\ref{eq:noisech}) describes a coupling between system and environment. We also assume that properties of channels and devices have been completely characterized beforehand and thus Alice and Bob, two parties performing the protocol, know in advance how local devices and channels coupled with environment behave accordingly.

\subsubsection*{The noise model} 

\begin{figure}[h!]
\centering\rule{0pt}{4pt}\par
%\scalebox{1}
{\includegraphics[width= 8cm]{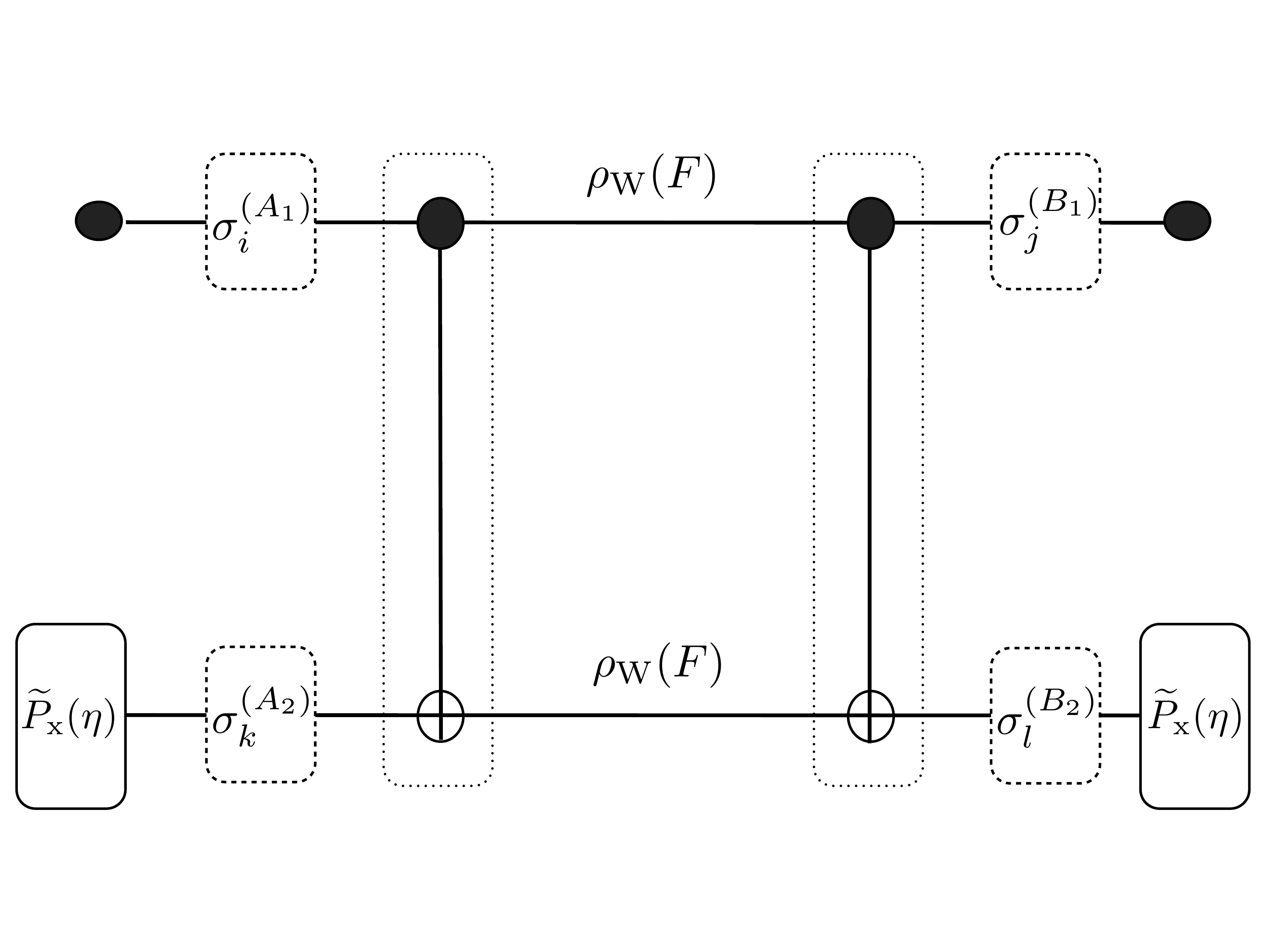}}
\vspace*{8pt}
\caption{\label{fig:protocol}
The distillation protocol takes two copies of Werner states in the four arms $A_1 B_1 A_2 B_2$, all of which can interact with environment locally. Noise in measurement can transferred to errors appearing in the second register $A_2 B_2$, see Eq. (\ref{eq:sim}). Errors appearing in the four locations are denoted by Pauli matrices, $\sigma_{i}^{(A_1)}$, $\sigma_{j}^{(B_1)}$, $\sigma_{k}^{(A_2)}$, and $\sigma_{l}^{(B_2)}$, and they happen with probabilities $C_{ijkl}$, see Eq. (\ref{eq:noch2}).  } %Later, in the post-selection step in the protocol, there can be imperfection in the projective measurement. Once measurement outcomes in the second register coincide each other, the state in the first register is accepted. }
\end{figure}

Having fixed notations and terminologies, we now introduce a general formalism of the distillation protocol with noisy operations in terms of noise effects on the distillation map. As we are interested in local noise, four locations where noise can happen locally are $A_1$, $A_2$, $B_1$, and $B_2$ in the bilateral CNOT operation $\E_{U}$ and the second register $A_2$ and $B_2$ in the projective measurement, see Fig. \ref{fig:protocol} for the full consideration. 

Let $\N_{L}^{(AB)}$ where $AB = A_1 B_1 A_2 B_2$ denote the noise map that describes local noise in the four locations as follows
\bea
\N_{L}^{(AB)} = \sum_{ijkl} p_{ijkl} ~\N_{q_i}^{(A_1)}  \otimes \N_{q_j}^{(B_1)}  \otimes \N_{q_k}^{(A_2)}  \otimes \N_{q_l}^{(B_2)}  \label{eq:noch}
\eea
where each noisy channel $\N_{q}$ can be found in Eq. (\ref{eq:noisech}). Let $\M_{L}^{(A_2B_2)}$ denote the noise map for the projective measurement. That is, noise effects on both operations lead to modifications on the original operations as follows,
\bea
\E_{U} \rightarrow \N_{L}^{(AB)} \circ \E_{U}~\mathrm{and}~ P^{(A_2B_2)} \rightarrow \M_{L}^{(A_2B_2)} \circ  P^{( A_2B_2)}. \nonumber 
\eea
Then, the two-way distillation map in Eq. (\ref{eq:emap}) under local noise can be described as
\bea
\widetilde{\E} (\cdot) =   p_{\mathrm{succ}}^{-1} \tr_{A_2 B_2} [\N_{L}^{(A  B )} \circ \E_{U}(\cdot)~ \M_{L}^{(A_2 B_2)} \circ P^{(A_2 B_2)}] \label{eq:nomap} 
\eea
where $p_{\mathrm{sucuss}}$ denotes the probability of accepting measurement outcomes in the second register,
\bea
p_{\mathrm{succ}}  =  \tr_{AB} [\N_{L}^{(A  B )} \circ \E_{U}(\cdot)~ \M_{L}^{(A_2 B_2)} \circ P^{(A_2 B_2)}]. \nonumber
\eea
In the above, the noise map $\M_{L}^{(A_2 B_2)}$ for measurement can be absorbed to the noise map for the channel and thus  equivalently dealt as noise in channels. This follows from the relation in Eq. (\ref{eq:sedu}) and is to be discussed in the next subsection.

To apply these formulations to analyzing disitllability of entangled states, let us further evaluate the noisy map. In particular, we express the noisy bilateral CNOT operation as follows.
\bea
&& \N_L \circ \E_{U} (\cdot)  \nonumber\\
&=&  \sum_{ijkl} p_{ijkl} ~\N_{q_i}^{(A_1)}  \otimes \N_{q_j}^{(B_1)}  \otimes \N_{q_k}^{(A_2)}  \otimes \N_{q_l}^{(B_2)} \circ \E_{U} \nonumber\\
&=&  (1-p) \E_{U} (\cdot) +  \nonumber \\
&& ~~~~~~~\sum_{ijkl} C_{ijkl} \sigma_{[ijkl]}^{(A_1 B_1 A_2 B_2)} \E_{U} (\cdot) \sigma_{[ijkl]}^{(A_1 B_1 A_2 B_2)} ~~\label{eq:noch2}
\eea
where $\sigma_{[ijkl]}^{(A_1 B_1 A_2 B_2)} = \sigma_{i}^{(A_1)} \otimes \sigma_{j}^{(B_1)} \otimes \sigma_{k}^{(A_2)} \otimes \sigma_{l}^{(B_2)}$ and it holds that $ \sum_{ijkl}C_{ijkl} =  p $ and $C_{ijkl}\geq 0$ for $i,j,k,l =I,X,Y,Z$, where $I$ means the identity operator, see also Fig. \ref{fig:protocol}. There are $4^4$ parameters $C_{ijkl} $ to describe the noisy channel in the above, corresponding to $4^4$ types of local noise: 
\bea
IIII~~ IIIX~~ IIIY~~  IIIZ ~~ IIXI ~~ IIXX ~~\cdots ~~   ZZZZ. \nonumber
\eea
A coefficient $C_{ijkl}$ shows the probability that local noise $\sigma_{i}$, $\sigma_{j}$, $\sigma_{k}$ and $\sigma_{l}$ appear in registers $A_1$, $B_1$, $A_2$, and $B_2$, respectively. 

Note that it is the distribution $\{ C_{ijkl}\}_{ijkl}$ that can be characterized from devices beforehand. Thus, we assume that $\{ C_{ijkl}\}_{ijkl}$ are known from given devices. The depolarization noise corresponds to the case when all $C_{ijkl}$'s are put equal, i.e. $C_{ijkl}=p/4^{4}$ for all $i,j,k,l=I,X,Y,Z$. For convenience, we write by $IJKL$ to denote a type of noise that appears with probability $C_{IJKL}$. 

% Later, we reproduce this noise model when considering general local noise in the distillation process. We are interested in how distributions of noise, or what types of errors, are related to the distillability.

%In this subsection, we employ the noise model introduced in the previous section, to the distillation protocol. This basically analyzes the formulae in Eq. (\ref{eq:nomap}). There, noisy channels are applied to both channels and measurements. This, however, can be simplified. We first show how one can simplify the consideration, and the proceed to the main analysis.

\subsection{Simplification: imperfect measurement  to noisy operations}

We show that consideration of noise in measurement can be transferred into noise in channels. That is, the consideration of noise on measurement $\M_{L}^{(A_2 B_2)}$ can be equivalently considered as noise on the bilateral operation $\N_{L}^{(AB)}$. 
This is based on the duality relation shown in Eq. (\ref{eq:sedu}). Recall the noisy map $\widetilde{\E}$ in Eq. (\ref{eq:nomap}) and it can be rewritten as,
\bea
\widetilde{\E} (\cdot) & = & \tr_{A_2 B_2} [\N_{L}^{(A  B )} \circ \E_{U}(\cdot)~ \N_{L}^{(A_2 B_2)} \circ P^{(A_2 B_2)}]  \nonumber\\
& = &  \tr_{A_2 B_2} [ \N_{L}^{(A_2 B_2)} \circ \N_{L}^{(A  B )} \circ \E_{U}(\cdot)~ ~ P^{(A_2 B_2)}] ~~\label{eq:sim}
\eea
where $\N_{L}^{(A_2 B_2)} = \sum_{st}r_{st} \N_{s}^{(A_2)} \otimes \N_{t}^{(B_2)}$ on the second register. Then, the composition of two local noise channels is again in the form of local noise channel $\N_{L}^{(AB)}$, i.e.
\bea
\N_{L}^{(A_2 B_2)} \circ \N_{L}^{(A  B )} \equiv \N_{L}^{(A  B )} \nonumber
\eea
with a new distribution of $C_{ijkl}$ in the expression of Eq. (\ref{eq:noch2}). This shows that, to classify noise effects from channels and measurement, it suffices to consider noise effects of measurement. Then, from noise effects of channels obtained, the noise effects from measurement can also be explained. %We will also show how this can be done, later on.
Hence, without loss of generality, we focus on analyzing the following noise model,
\bea
\widetilde{\E} (\cdot) & = &  p_{\mathrm{succ}}^{-1} \tr_{A_2 B_2} [\N_{L}^{(A  B )} \circ \E_{U}(\cdot)~  P^{(A_2 B_2)}]  ~\mathrm{with} \label{eq:nomap2} \nonumber\\
p_{\mathrm{succ}} & = & \tr_{A B} [\N_{L}^{(A  B )} \circ \E_{U}(\cdot)~  P^{(A_2 B_2)}]  \label{eq:noisymap}
\eea
where the noisy channel can be found in Eq. (\ref{eq:noch2}) with ideal measurement.

\subsection{Entanglement distillability under local noise}

%After all, i.e. having passed local noise appearing in channels and measurement, our interest lies at the possibility of distilling entanglement. This can be summarized a simple form, in terms of the resulting singlet fraction, or equivalently the increment of the singlet fraction by an iteration of the entanglement distillation protocol.

From the simplification shown in the above, it suffices for us to consider the singlet fidelity increment under the noisy operations in Eq. (\ref{eq:noisymap}).
%in the following,
%\bea
%\widetilde{\E} (\rho_{\W}^{\otimes 2} (F)) & = & \frac{1}{p_{\mathrm{succ}}} \tr_{A_2 B_2} ~[ P^{(A_2 B_2)}  ~ \widetilde{\E}_{U} (\rho_{\W}^{\otimes 2} (F))],~~ \label{eq:nres} \\
% \mathrm{where} ~~ \widetilde{\E}_{U} & = & \N_{L}^{(A  B )} \circ \E_{U} \nonumber \\
% \mathrm{and} ~~  p_{\mathrm{succ}} & = &  \tr_{A B} ~[ P^{(A_2 B_2)}  ~ \widetilde{\E}_{U} (\rho_{\W}^{\otimes 2} (F))]. \nonumber
%\eea
With the noisy operation, the singlet fidelity increment is given by
\bea
\delta \widetilde{F} = \widetilde{F}^{'} - F,~~\mathrm{where} ~ \widetilde{F}^{'} = \langle \phi^{+}  | \widetilde{\E} (\rho_{\W}^{\otimes 2} (F))  | \phi^{+}\rangle\label{eq:sinfid}
\eea
where $\widetilde{F}^{'}$ denotes the fidelity of a resulting state from the noisy distillation map. We also recall that by the protocol entanglement increases if $\delta \widetilde{F} >0$. We analyse the disitllability with the description in Eq. (\ref{eq:noch2}), evaluating the singlet fidelity,
\bea
\widetilde{F}' & =&  (1-p)F' + \sum_{ijkl} C_{ijkl} F'_{ijkl},~\mathrm{where}~~  \label{eq:nofde} \\
F' &= & p_{\mathrm{succ}}^{-1}  \langle \phi^{+} |  \tr_{A_2 B_2}   P^{(A_2 B_2)}   \E_U ( \rho_{W}^{\otimes 2} (F))      |\phi^{+} \rangle \nonumber\\
F'_{ijkl} & = &  p_{\mathrm{succ}}^{-1}  \langle \phi^{+} |   \tr_{A_2 B_2}    P^{(A_2 B_2)}  \nonumber \\
&&[  \sigma_{[ijkl]}^{(A_1 B_1 A_2 B_2)}   \E_{U} ( \rho_{W}^{\otimes 2} (F))     \sigma_{[ijkl]}^{(A_1 B_1 A_2 B_2)} ] | \phi^{+} \rangle ~~ ~~\label{eq:fijkl}
\eea
and $F'$ correspond to the singlet fidelity of the resulting state when noise is not present in the distillation protocol. %It is remarkable in the above that, the singlet fraction can be finally described in a linear expression.

From the relation in Eq. (\ref{eq:nofde}), the distillability condition can be shown in terms of $F_{ijkl}^{'}$: entanglement can be distilled if the singlet fidelity increment is positive:
\bea
0 < \delta \widetilde{F}' & = & \widetilde{F}' - F  \nonumber \\
& = & (1-p)F' - F + \sum_{ijkl} C_{ijkl} F'_{ijkl} \nonumber \\
%& = & (1-p)\delta F - pF +\sum_{ijkl} C_{ijkl} F'_{ijkl} \nonumber \\
%\equiv%\leftrightarrow 
\mathrm{i.e.,}&& (1-p) \delta F + \sum_{ijkl} C_{ijkl} F'_{ijkl} > p F \label{eq:discon}
\eea
where $\delta F$ is the singlet fidelity  increment in Eq. (\ref{eq:fincre}) when noise is not present in the operations of the entanglement distillation protocol. To obtain the distillability condition, it only remains to consider $\{ C_{ijkl}\}_{ijkl}$ and $\{ F_{ijkl}^{'} \}_{ijkl}$. Recall that parameters $\{C_{ijkl}\}_{ijkl}$ show distribution of types of local noise from properties of measurement devices in experiment: they are thus given in experiment.

\section{Analysis of the Distillability }

%In this section, we analyze the entanglement distillability under general local noise. In the condition Eq. (\ref{eq:discon}), the distillability condition depends on the distribution $C_{ijkl}$ of fidelities $F_{ijkl}$, where fidelities $F_{ijkl}$  are functions of the singlet fidelity $F$ of initially given Werner states according to the two-way distillation map, see Eq. (\ref{eq:fijkl}). In a realistic setting, the distribution $\{ C_{ijkl}\}_{ijkl}$ can be observed from characterisation of measurement devices: although it is not possible to know which error certainly happens when the protocol runs, it is possible to have in advance probabilistic behaviour of measurement devices from the device characterisation. 

%In the following, we consider fidelities $F_{ijkl}$ and, by comparing them, show equivalence classes of the fidelities or equivalently types of errors. To be explicit, 
%As it has been mentioned in the above, there are $4^4$ types of noise. 

In this section, we find explicit expressions of $F_{ijkl}^{'}$ in Eq. (\ref{eq:discon}) and analyse how they are related to the distillability condition. As it has been mentioned in the above, there are $4^4$ types of errors in which, however, we show that they do not always give distinct effects to the distillation protocol. In fact, there are only four distinct types of errors, i.e. equivalence classes of types of noise. 

For instance, suppose that for two types of noise, $ijkl$ and $abcd$ the resulting singlet fidelities are equal. This means from Eq. (\ref{eq:nofde})
\bea
\widetilde{F}' & = & (1-p) F' + p F'_{ijkl} \nonumber \\
& = & (1-p) F' + p F_{abcd}^{'}. \nonumber
\eea
where it has been used that $C_{ijkl} = C_{abcd} =p$.  It is clear that we have, $F'_{ijkl} = F'_{abcd}$. This shows that two kinds of noise have the same effect to the distillation protocol: more precisely, they are equivalent with respect to the distillability condition. We therefore consider two types of noise $ijkl$ and $abcd$ equivalent:
\bea
ijkl \sim abcd ~~\mathrm{if} ~~F_{ijkl}^{'} = F_{abcd}^{'}. \nonumber
\eea
We call a set of equivalent types of noise as Error Class. In the following, we show that there are in fact only four equivalence classes among the $4^4$ types of errors. Note that once types of noise are in the same equivalent class, any combinations of them are also in the same class, i.e. for $ ijkl\sim abcd$, 
\bea
\widetilde{F}' & = & (1-p) F' + C_{ijkl} F'_{ijkl} + C_{abcd} F'_{abcd} \nonumber \\
& = & (1-p) F' + p F'_{ijkl} \nonumber 
\eea
for all $C_{ijkl} + C_{abcd}=p$ and $C_{ijkl}\geq 0$ and $C_{abcd}\geq 0$.

%Relevant parameters to the distillability correspond to the range of fidelity in which entanglement increases by the protocol, that is, $F_{\min}$ and $F_{\max}$ such that between them we have $\delta \widetilde{F} > 0$. In the ideal case of the distillation protocol, $F_{\min} = 1/2$  and $F_{\max} =1$. To find two parameters $F_{\min}$ and $F_{\max}$, we follow the following steps.

%Entanglement distillation when local noise is present in the channel is shown, in particular, for $p=0.05$ in Eq. (\ref{eq:noismodel}). Note that the projective measurement is supposed to be ideal, i.e. $\eta_{\x}^{A_2} = \eta_{\x^{B_2}}=1$ for $\x=0,1$ in Eq. (\ref{eq:noproj}). Given $p=0.05$, different distributions of $C_{ijkl}$ where $(i,j,k,l=0,1,2,3)$ are considered to find how they are related to the increment of the singlet fidelity by an iteration of the protocol. For this purpose, we take the parameterization in Eq. (\ref{eq:cpara}) that takes one of $256$ parameters as the largest, and in particular choose $g=1$. Interestingly, from $256$ cases of noise distributions, resulting curves of the increment of the singlet fidelity are characterized into only $4$ cases, as $(I)$, $(M)$ $(C_1)$ and $(C_2)$, see also the main text. They are denoted by Error Classes. The class (D) denotes the case of depolarization channel, i.e. $C_{ijkl} = p/4^4$ which equals to Eq. (\ref{eq:deo}), and can also explained by $g\rightarrow \infty$ in Eq. (\ref{eq:cpara}). Note that, $F_{\max}=1$ for class $M$, and the distillability parameters $F_{\min}$ and $F_{\max}$ are compared.

\subsection{Equivalence classes among types of error}

\begin{figure}[h!]
\centering\rule{0pt}{4pt}
%\scalebox{1}
{\includegraphics[width= 9cm]{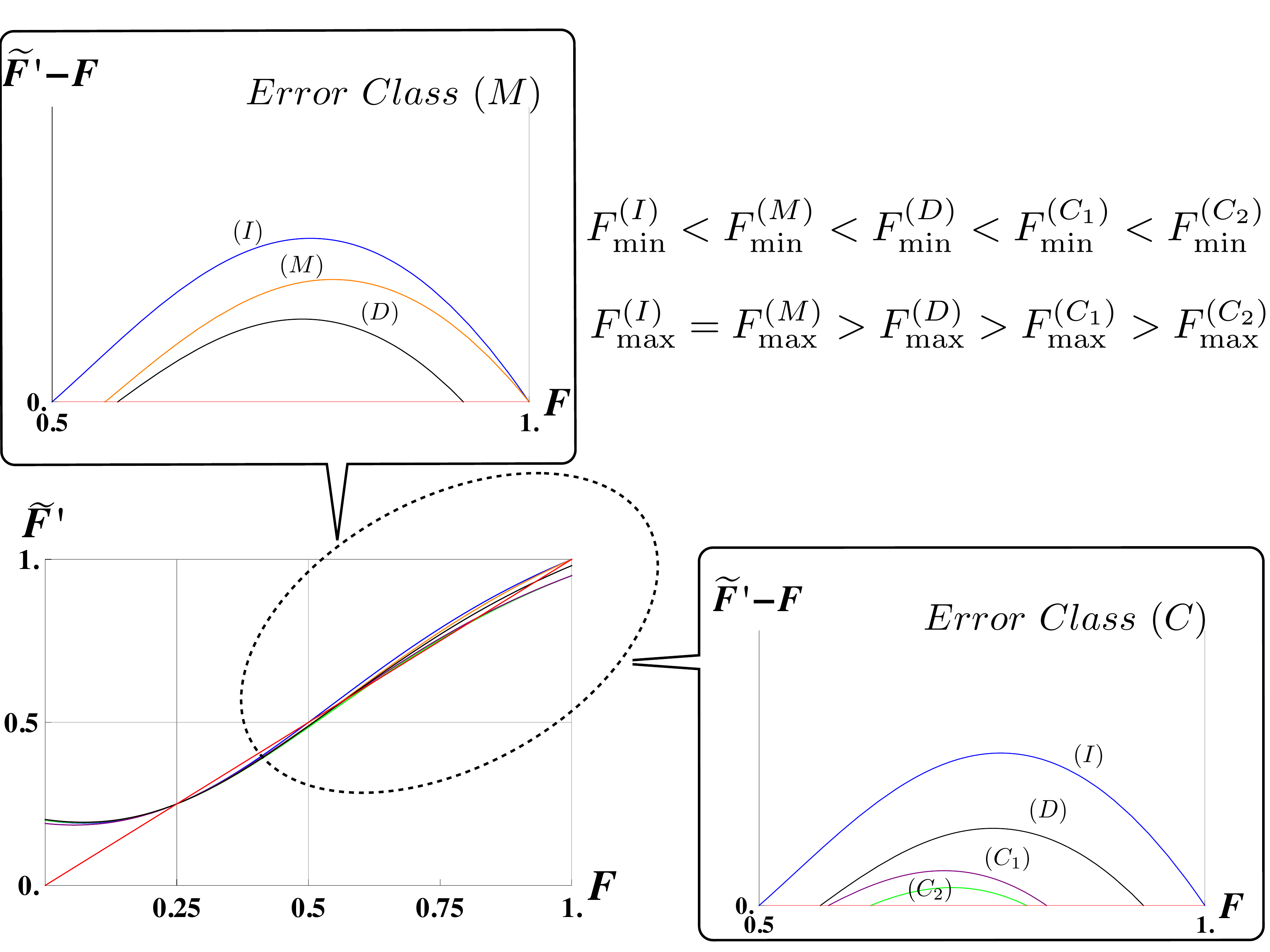}}
\vspace*{8pt}
\caption{\label{fig:fig6chno}
The distillation protocol under general noise runs for $4^4$ types of local noise. To find if two types $IJKL$ and $I'J'K'L'$ are equal or not, we put $C_{IJKL} = C_{I'J'K'L'} = p$ in Eq. (\ref{eq:nofde}) and compare the fidelity increment $\delta \widetilde{F}^{'}$. It turns out that there are four equivalence classes in the $4^4$ types of noise, called (I), (M), $(C_1)$, $(C_2)$. The curve $(D)$ means a random noise described by the depolarization channel, and corresponds to the case when all types of noise appear with equal probability $C_{ijkl} = 1/4^4$. Relevant parameters in the curve are $[F_{\min},~ F_{\max}]$: quantum states within the range are distillable. The class $(I)$ consists of the type $IIII$ that means the ideal case without any error, and show that all types in $(I)$ do not affect the distillation protocol. }
\end{figure}

To find if a type of noise is equivalent to another with respect to distillability of entanglement, we make the following analysis. Given a type of noise $IJKL$, we simplify the description in Eq. (\ref{eq:noch2}) such that 
\bea
\widetilde{\E} (\cdot) = (1-p) \E (\cdot) + p ~\sigma_{IJKL}^{(A_1A_2B_1B_2)} \E(\cdot)  \sigma_{IJKL}^{(A_1A_2B_1B_2)} \nonumber
\eea
and then compute the corresponding fidelity increment in Eq. (\ref{eq:discon}). This repeats for all $4^4$ types of noise, and the numerical results are presented in Fig. (\ref{fig:fig6chno}). Remarkably, there are only four distinct curves $\delta \widetilde{F}^{'} = \widetilde{F}^{'} - F$, see also Eq. (\ref{eq:discon}). This means that there are only four distinct types of noise among all of $4^4$ ones. We then collect equivalent types of noise, that define equivalence classes of types of noise.

To describe them we write types of errors as follows. Among four Pauli matrices as $I$, $X$, $Y$, and $Z$, let $P$ denote one of phase-operators, $I$ or $Z$, and $B$ be one of bit-operators, $X$ or $Y$:
\bea
P \in \{ I,Z \},~~ \mathrm{and}~~  B \in \{ X,Y \}. \label{eq:pb}
\eea
For instance, we write by $PP$ to denote all combinations of phase-operations and $PB$ the four possibilities of phase and bit operations:
\bea
PP:= \{ P\} \otimes \{ P\} & = &  \{ II,IZ,ZI,ZZ\} \nonumber \\ 
PB:= \{ P \} \otimes \{ B\} & = & \{ IX,IY,ZX,ZY \}. \nonumber
\eea 
Among all of $4^4$ types of local noise, we show that there are four distinct classes only. We call them Error Class.
\\
%We write, by $IJKL$, when a fixed $C_{IJKL}$ ($=p$) is only non-zero. For instance, by saying that we consider the singlet fidelity caused by noise in the form $IIIZ$, the parameter $C_{IIIZ}$ is the noise most dominant among others.

\textbf{Error Class (I). }
The class (I) consists of the case $IIII$, that corresponds to the ideal case in Eq. (\ref{eq:emap}). Therefore, this class collects those errors which do not affect the distillation protocol at all. These are summarized:
\bea
(I): ~~ \{II,XX,YY,ZZ  \}^{(A_1 , B_1)}  \otimes \{ PP,BB \}^{(A_2 , B_2)}.   \label{eq:ecIIII}
\eea
$P$ and $B$ in the above can be found in Eq. (\ref{eq:pb}). For examples, $XXZZ$, $XXII$, $XXIZ$, $XXXX$, $XXXY$, etc. are in this class, and there are $32$ instances. If noise in this class happens in the distillation protocol the distillability condition remains the same, i.e. all entangled two-qubit states can be distilled. Let us summarize this by, 
\bea
(I): ~~[ F_{\min}^{(I)}, ~F_{\max}^{(I)} ] = [1/2,~1]. \nonumber 
\eea
It is also worth to observe that noise appearing in the second register is either $PP$ or $BB$, i.e., an identical type in both parties. \\

\textbf{Error Class (M). }
The next class, called Error Class (M), can be summarized as follows
\bea
(M): ~~ \{ PP, PB, BP, BB \}^{(A_1 , B_1)}  \otimes \{ PB,BP \}^{(A_2 , B_2)}.   \label{eq:ecM}
\eea
For instance, $ZZZX$, $ZZZY$, $ZXXZ$, etc. are in this class, and in this way we have $128$ instances. This class is denoted by $(M)$ since equivalent types of errors under noise in channels are, as we will show later, to be further classified and show distinct distillation curves by noise in measurement.

It is worth to observe that the maximally attainable singlet fidelity is equal to the unit, i.e. $F_{\max}^{(M)}=1$. However, unlike Class $(I)$, errors in this class are critical as $F_{\min}^{(M)}>1/2$, that is, depending on the noise, some weakly entangled states cannot be distilled. Compared to the case of depolarization noise considered in Ref. \cite{ref:Dur}, the distillation protocol is more robust against this class $(M)$. Denoted by $[F_{\min}^{(D)} , F_{\max}^{(D)}]$ the range in which entanglement increases by the protocol when a depolarization noise is present, it holds that 
\bea
[F_{\min}^{(D)} , ~F_{\max}^{(D)}]  \subset [F_{\min}^{(M)} , ~F_{\max}^{(M)}] = [F_{\min}^{(M)}, ~1]  \subset [1/2,~1]. \label{eq:result1}
\eea
That is, more of entangled states can be distilled than the case of depolarization noise and maximally entangled states can be distilled. see also Fig. \ref{fig:fig6chno}.
\\

\textbf{Error Class ($C_1$) and ($C_2$). }
We finally collect two classes, denoted by ($C_1$) and $(C_2)$,
\bea
(C_1) &:& \{ IZ, ZI, XY, YX  \}^{(A_1 , B_1)}  \otimes \{ PP,BB \}^{(A_2 , B_2)}, ~~~~~~\label{eq:ecC1}\\
(C_2) &: & \{PB, BP \}^{(A_1 , B_1)}  \otimes \{ PP,BB \}^{(A_2 , B_2)}.  ~~~~~~\label{eq:ecC2}
\eea
We call them as $C_1$ of $32$ instances,  and $C_2$ of $64$ instances, respectively, since these classes show distinct distillability conditions due to noise in channel. Let us summarize as follows,
\bea
 [F_{\min}^{(C_2)} , ~F_{\max}^{(C_2)}]  \subset [F_{\min}^{(C_1)} , ~F_{\max}^{(C_1)}]  \subset  [F_{\min}^{(D)} , ~F_{\max}^{(D)}]. \label{eq:result2}
\eea
That is, these types of noise $C_1$ and $C_2$ are more critical to the protocol than in the case of depolarization channel. It holds that two thresholds are strictly weaker than those of the depolarization, i.e.  
\bea
~~F_{\min}^{(C_2)} > F_{\min}^{(C_1)} > F_{\min}^{(D)},~\mathrm{and}~ F_{\max}^{(C_2)} < F_{\max}^{(C_1)}  < F_{\max}^{(D)}.\nonumber
\eea

\subsection{Analytic expression }

We have shown that among $4^4$ types of noise, there are only four distinct ones. From Eq. (\ref{eq:discon}), this means that there only four distinct analytic expression for fidelities $F_{ijkl}^{'}$. We write these fidelities by $F_{K}^{'}$ as follows, where $K$ means types of error in each Error Class - $(I)$, $(M)$, $(C_1)$, and $(C_2)$:
 %Thus, to have analytic formulas for these four fidelities, $F_{ijkl}^{(K)}$, we have analytic expressions for the singlet fidelity $\widetilde{F}'$ for all types of noise.\
\bea
\mathrm{Error ~Class~ (I)}: && F_{I}^{'} =  p_{\mathrm{succ}}^{-1} \frac{ 1 - 2F +10F^2 }{9 },  \nonumber  \\
\mathrm{Error ~Class ~(M)}: && F_{M}^{'} =  p_{\mathrm{succ}}^{-1} \frac{ 1 + F -2 F^2 }{9 },  \nonumber  \\
\mathrm{Error~ Class ~(C_1) } && F_{C_1}^{'} = p_{\mathrm{succ}}^{-1} \frac{ 2(1-F) F }{ 3 }, \nonumber \\
\mathrm{Error ~Class ~(C_2)}: && F_{C_2}^{'} = p_{\mathrm{succ}}^{-1} \frac{ 2 (1-F)^2 }{9 }, \nonumber 
\eea
where the success probability is given by
\bea
p_{\mathrm{succ}} = \frac{1}{9} [ (5-4F+8F^2) - \frac{p}{2} (1-4F)^2] \nonumber
\eea
and $p$ denotes the probability that types of noise may happen, see Eq. (\ref{eq:noch2})

%Recall that the resulting singlet fidelity is expressed in Eq. (\ref{eq:nofde}), and there are only four types of errors. This means that, there only four types in $F_{ijkl}$ depending on Error Classes. We write them by $F_{ijkl}^{K}$ where $K$ means Error Class - $(I)$, $(M)$, $(C_1)$, and $(C_2)$. Thus, to have analytic formulas for these four fidelities, $F_{ijkl}^{(K)}$, we have analytic expressions for the singlet fidelity $\widetilde{F}'$ for all types of noise.

\subsection{Reproducing the depolarization}

The depolarization noise can be described as the case when all types of noise appear with the same probability, i.e. $C_{ijkl}=p/4^4$ in Eq. (\ref{eq:noch2}). According to the classification in the above, the depolarization case can be reproduced as follows, in terms of Eq. (\ref{eq:nofde}),
\bea
\widetilde{F}'^{(D)} & = & (1-p) F' + \frac{p}{4^4} \sum_{ijkl} F_{ijkl}^{'} \nonumber\\
& = & (1-p) F' + \frac{p}{4^4} (32 F_{I }^{' } + 128 F_{ M}^{' } + 32 F_{C_1}^{'} \nonumber \\
&& + 64 F_{ C_2 }^{' }), \nonumber
\eea
where we have used the cardinality of equivalence classes. This can be computed as,
\bea
\widetilde{F}'^{(D)} & = & (1-p) F' + \frac{p}{8} F_{I }^{ ' } + \frac{p}{2 } F_{M }^{ ' } + \frac{p}{8}F_{C_1 }^{' } + \frac{p}{4} F_{C_2 }^{ ' }, \nonumber \\
& = & (1-\frac{7}{8}p) F' +  \frac{p}{2 } F_{M }^{' } + \frac{p}{8} F_{ C_1 }^{ ' } + \frac{p}{4} F_{C_2 }^{ ' }, \nonumber
\eea
since the class $(I)$ does not affect to distillability. Thus, the depolarization noise is reproduced. 

\begin{figure}[h!]
\centering\rule{0pt}{4pt}\par
%\scalebox{1}
{\includegraphics[width= 8.8cm]{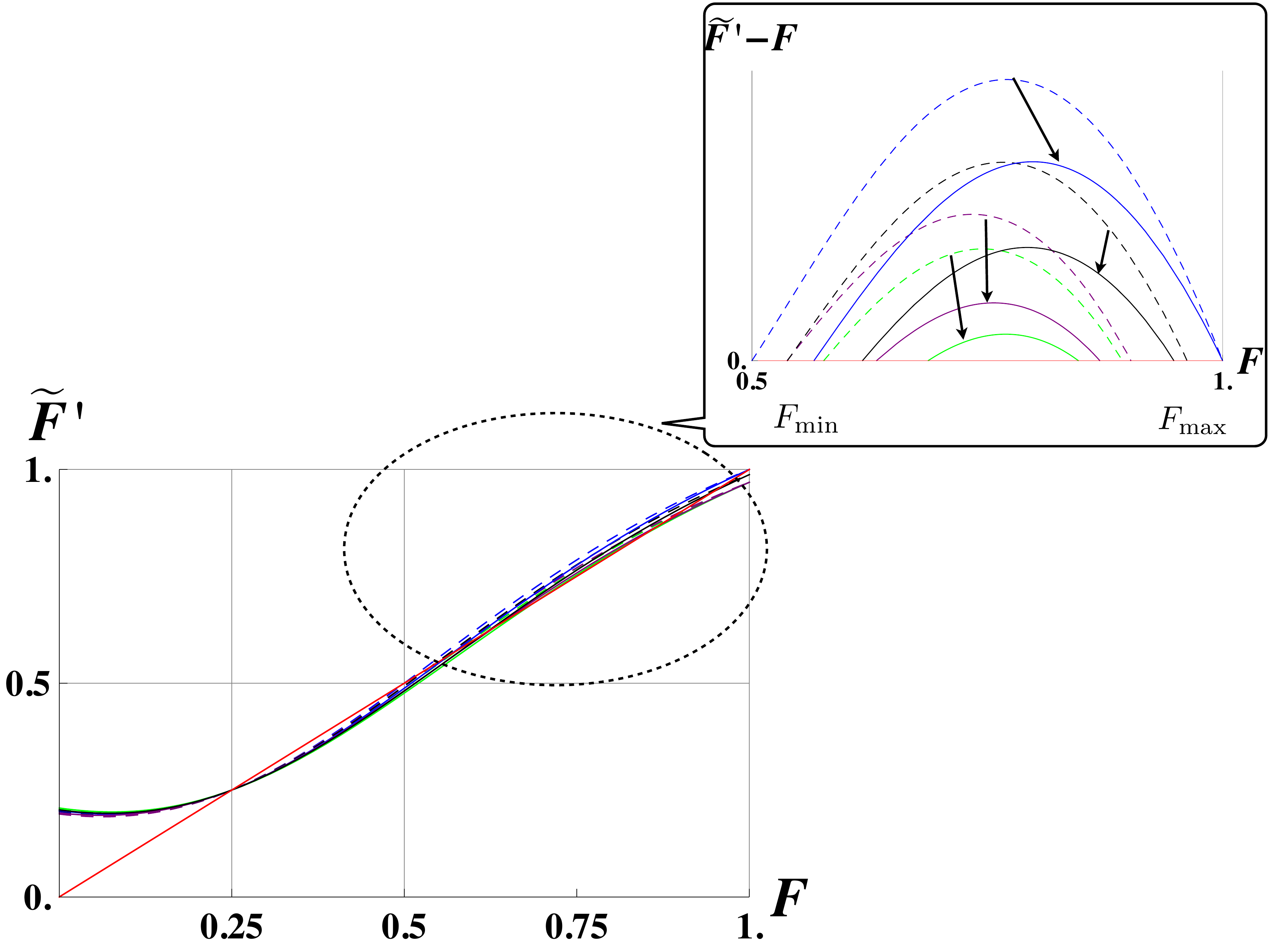}}
\vspace*{8pt}
\caption{\label{fig:fig8clC}
The distillation curves $\delta \widetilde{F}^{'} = \widetilde{F}^{'} -F$ are shown when noise happens in measurement for classes $(I)$, $(C_1)$, $(C_2)$, and the depolarization $(D)$. Relations of these types of noise can be found in Eqs. (\ref{eq:result1}) and (\ref{eq:result2}). It is shown that the distillable areas are reduced for all cases.}
\end{figure}

\subsection{ Local noise in measurement}

\begin{figure}[h!]
\centering\rule{0pt}{4pt}\par
%\scalebox{1}
{\includegraphics[width= 8.8cm]{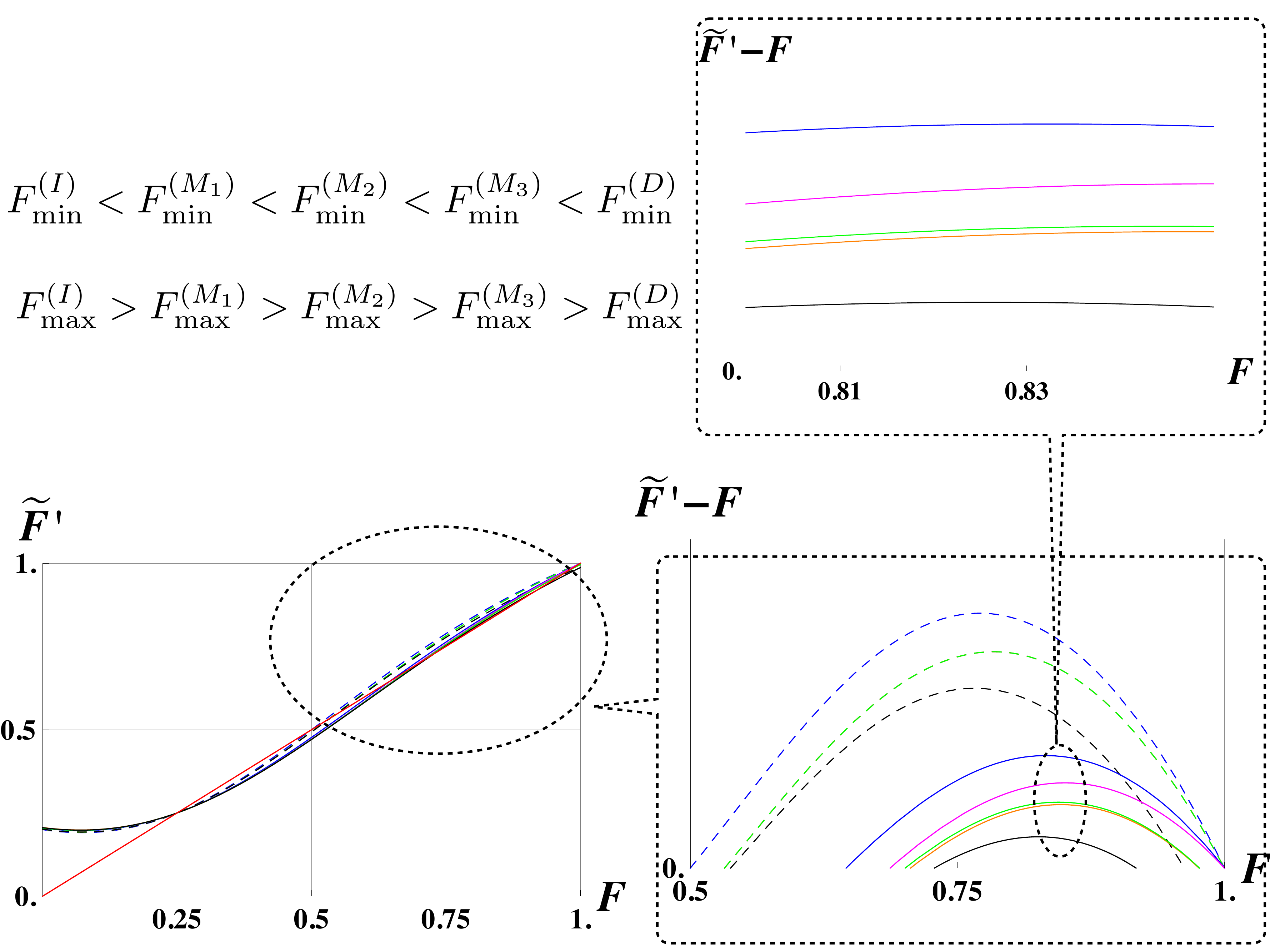}}
\vspace*{8pt}
\caption{\label{fig:fig8clM}
The distillation curves $\delta \widetilde{F}^{'} = \widetilde{F}^{'} - F$ are shown when noise happens in measurement for classes $(I)$, $(M)$, and the depolarization $(D)$. It is shown that the distillable areas are reduced for all cases. Relations of these types of noise can be found in Eqs. (\ref{eq:result1}) and (\ref{eq:result2}). It is found that the class $(M)$ is split into three distinct types, denoted by $M_1$, $M_2$, and $M_3$. }
\end{figure}

In addition, let us consider local noise appearing in measurement. As we have discussed and shown in Eq. (\ref{eq:noisymap}), noise in measurement can be equivalently considered as noise in operations, and hence does not introduce a new type of noise other than what we have shown so far. We are here interested in how effects of noise in measurement modify distillability as well as efficiency of distillation. Let us present numerical simulation on this.

Recall that the projective measurement is given by, $P_{x} = |\x \rangle \langle \x |$ for $\x=0,1$. Then, an imperfect measurement is described by,
\bea
\widetilde{P}_{\x} (\eta) = (1-\eta) | \x\rangle \langle \x| +\eta |\x+1\rangle \langle \x+1|, \nonumber
\eea
with noisy parameter $\eta$ where the addition is computed modulo $2$. Using the formulation of the noise map in Eq. (\ref{eq:noisech}), noise appearing in measurement can be equivalently considered as cases that only bit-flip errors happen in operations. The results are shown numerically Figs. (\ref{fig:fig8clC}) and (\ref{fig:fig8clM}).

%That is, denoted by $\N(X)$, the noisy measurement can be expressed as,
%\bea
%\widetilde{P}_{x}(\eta) = \N(X) \circ P_{x},~~\N(X)(\cdot ) = (1-\eta)(\cdot ) + \eta \sigma_{X} (\cdot ) \sigma_{X}. \nonumber
%\eea
%Using the duality relation in Eq. (\ref{eq:sedu}), this noise channel can be alternatively transferred to the channel, as it is shown in Eq. (\ref{eq:sim}). The noise transferred to channel from measurement is only concerned to the bit-flip error to the second pair $A_2$ and $B_2$. Finally, let us show numerical results when both errors are considered, see Figs. (\ref{fig:fig8clC}) and (\ref{fig:fig8clM})

%[LATER, THIS PART SHOULD BE MORE ANALYZED]
%Bit-flip errors in the second copy of either or both parties corresponds to modifying Error Classes. For instance, Error Classes $C_1$ or $C_2$ that have $\{PP,BB \}$ in the second copy can be transformed to $PB,BP,PP,BB $, in which they can become in the Error Class $(M)$ if the second copy has $PB$ or $BP$. When bit-flip noise happens for the class $(M)$, the second copy can be transformed to $BB,BP,PB,PP$. When transformed to $PP$ or $BB$, these do not correspond to

%\section{Interpretation: Usefulness of Local Random Operations}

%\section{Cost reduction in long-distance communication}

\section{Conclusion}

Distilling entanglement is a fundamental task in quantum information processing. Given quantum states, deciding if they are distillable or undistillable is theoretically challenging in general, and so far it has been known from the two-way distillation protocol that all two-qubit entangled states are distillable. Then, distillability for two-qubit entangled states once quantum operations in the protocol are noisy is of practical interest since systems often interact with and are consequently coupled to local environment. 

In this work, we have considered entanglement distillation in the realistic setting where quantum operations are noisy due to interaction between systems and environment. We have assumed that specifications of devices applied to entanglement distillation are characterized in advance such that probability distributions of different types of noise are known beforehand and can be exploited when analyzing distillability. This is also realistic since in a laboratory one is often not in cases of knowing nothing about properties of devices but, in fact, has $a~ priori$ information about how often a type of noise would appear in experiment. Hence, with such $a~priori$ information at hands, one does not have to necessarily to reduce the consideration to a random noise described by a depolarization channel. 

In a single round of the two-way distillation protocol, two copies of Werner states are shared by and located in the four arms, denoted by $A_1B_1 A_2 B_2$, of Alice and Bob. These are the four locations that systems are coupled to environment locally. We have shown that among all possible $4^4$ types of noise there are only four distinct ones in terms of distillability conditions: namely, class (I) having $32$ instances not affecting to distillability, class (M) of $128$ instances less critical than the depolarization, class $(C_1)$ of $32$ instances, and class $(C_2)$ of $64$ instances. One can also consider noise effects in measurement in the second register $A_2 B_2$. We have shown that this can be equivalently considered as noise in quantum operations in arms $A_2 B_2$. We have presented these results in a general formalism of the two-way distillation protocol with noisy operations. 

Our results find that the distillation protocol is more robust to the two types of noise classes (I) and (M), of $160$ instances overall out of $4^4$ ones, than the depolarization case. This shows the usefulness of  $a~priori$ information about noise properties of devices applied to entanglement distillation. One may also envisage that applying to entanglement distillation between quantum repeaters for long-distance communication, these results can be exploited to extended the communication distance.

\section*{Acknowledgment}
This work is supported by Institute for Information \& communications Technology Promotion(IITP) grant funded by the Korea government(MSIP) (No.R0190-16-2028, PSQKD), the research fund of Hanyang University (HY-2015-259), and the National Research Foundation of Korea (NRF-2010-0025620). 

%\ifCLASSOPTIONcaptionsoff
%  \newpage
%\fi



%\bibliographystyle{IEEEtran}

\begin{thebibliography}{1}
  
\bibitem{ref:IBM} C.H. Bennett, G. Brassard, S. Popescu, B. Schumacher, J. Smolin, and W.K. Wootters, \textit{Purification of Noisy Entanglement and Faithful Teleportation via Noisy Channels}, Phys. Rev. Lett. \textbf{76}, (1996) 722.

\bibitem{ref:Oxford} D. Deutsch, A. Ekert, R. Jozsa, C. Macchiavello, S. Popescu, and A. Sanpera, \textit{Quantum Privacy Amplification and the Security of Quantum Cryptography over Noisy Channels}, Phys. Rev. Lett. \textbf{77}, (1996) 2818.

\bibitem{ref:3} C. H. Bennett, D. P. DiVincenzo, J. A. Smolin, W. K. Wootters, \textit{Mixed State Entanglement and Quantum Error Correction}, Phys. Rev. A {\textbf 54} (1996) 3824-3851.

\bibitem{ref:4} C. H. Bennett, H. J. Bernstein, S. Popescu, B. Schumacher, \textit{Concentrating Partial Entanglement by Local Operations}, Phys. Rev. A {\textbf 53} (1996) 2046-2052.

\bibitem{ref:durcirac} W. D\"ur, J. I. Cirac, M. Lewenstein, and D. Bruss, \textit{Distillability and partial transposition in bipartite systems}, Phys. Rev. A \textbf{61}, (2000) 062313.

\bibitem{ref:Briegel} H.-J. Briegel, W. D\"ur, J.I. Cirac, and P. Zoller, \textit{Quantum Repeaters: The Role of Imperfect Local Operations in Quantum Communication}, Phys. Rev. Lett. \textbf{81}, (1998) 5932.

\bibitem{ref:Dur} W. D\"ur, H.-J. Briegel, J.I. Cirac, and P. Zoller, \textit{Quantum repeaters based on entanglement purification}, Phys. Rev. A. \textbf{59}, (1999) 169.

\bibitem{ref:2design} C. Dankert, R. Cleve, J. Emerson, and E. Livine, Phys. Rev. A \textbf{80}, (2009) 012304. 

\bibitem{ref:noclone} W. K. Wootters and W. H. Zurek, \textit{A Single Quantum Cannot be Cloned}, Nature, {\textbf 299}, (1982) 802-803.

\bibitem{ref:mau} U. Maurer, \textit{Secret key agreement by public discussion from common information}, IEEE Transactions on Information Theory, Vol. \textbf{39}, 3, (1993) 733-742.

\bibitem{ref:high} P. Horodecki and R. Horodecki, \textit{Distillation and Bound entanglement}, Quantum Information and Computation, Vol. 1, No.1, (2001) 45-75. 

\bibitem{ref:geza} G. Giedke, H. J. Briegel, J. I. Cirac, and P. Zoller, \textit{Lower bounds for attainable fidelities in entanglement purification}, Phys. Rev. A, {\textbf 59}, (1999) 2641. 

\bibitem{ref:werner} R. F. Werner, \textit{Quantum states with Einstein-Podolsky-Rosen correlations admitting a hidden-variable model}, Phys. Rev. A. \textbf{40}, (1989) 4277. 


\bibitem{ref:PereHoro} A. Peres, \textit{Separability Criterion for Density Matrices}, Phys. Rev. Lett. \textbf{77}, (1996) 1413; M. Horodecki, P. Horodecki, R. Horodecki, \textit{Separability of Mixed States: Necessary and Sufficient Conditions}, Phys. Lett. A \textbf{223}, (1996) 1.


\end{thebibliography}


%\vfill


\end{document}